# Purely electrical nature of ball lightning (BL), its elementary equations, calculated parameters and conditions of possible BL experimental generation [1, 2, 3]

V. N. Soshnikov

Plasma Physics Dept.,
All-Russian Institute of Scientific and Technical Information
of the Russian Academy of Sciences
(VINITI, Usievitcha 20, 125315 Moscow, Russia)

**Abstract**

The nature of ball lightning (BL) is pure electric and can be described by simple equations following to elementary considerations of equality of translational accelerations and velocities of BL ions and electrons, a spherical-like dipole BL as a whole and balance of the energy influx of atmospheric electricity and radiation losses. From these equations follows a linear relationship between the size of BL and the tension of the atmospheric field E. A typical size of the fireball (FB) r ~ (5 − 10) cm corresponds to the calculated electron temperature $T_e$ ~ 8000 K at a pressure p = 1 at with a horizontal component of the electric field E is a few kV/cm. I estimate the energy of BL and characterize the conditions of its possible experimental generation. The estimation is given of the surface tension of BL. The possibility of the "hot" and the most realistic thermodynamic non-equilibrium "cold" BL is discussed. Here we presented preliminary evaluations preceding the more detailed work in Arxiv.org [11].

## 1. Introduction

The impossibility until now to obtain BL experimentally, variety of its manifestations, enormous quantity of "evidences" up to the semi-fantastic communications about the penetration BL through the intact glass, the sizes BL up to ~1.2 m and more, its tracking aircraft and about other numerous mysterious phenomena connected with it, they entailed many uncommon hypotheses, for example, about the appearance of BL and its prolonged existence due to the radioactive particles inside it, idea about the fact that BL is the phase state of dense nonideal plasma with the temperature $T$ ~400 K; that BL is the state of combustion in air of the hard cluster structure, which was being formed as a result of the impact of linear lightning into the metallic beam or the construction with the subsequent thin dispersion of material, hypothesis of Rydberg atoms with high BL surface tension, finally, the well known hypothesis that BL is generated and exists in antinodes of standing electromagnetic radio-wave, and also other hypotheses (numerous evidences of eyewitnesses, hypotheses and references to the written sources can be found in Internet and also in numerous descriptions of BL, for example, in [1] and complementary sequence [2], [3], [4] and others).

---

[1] Reexamined and completed.

[2] E-mail: vikt3363@yandex.ru

[3] In this paper, there is no clear indication of polarization of BL plasma. Account for this is made in subsequent work and can be made here by simply replacing the notation of the electric field $E_{env}$ by $E_{eff}$ in BL plasma, and with $E_{env}$ to denote the value of the real external (atmospheric) field $E_{env} = E_{eff}/\gamma$ where $0 < \gamma \leq 1$ (usually $\gamma \ll 1$, and $E_{env}$ is up to values of the order few kV/cm). As one can assume, factor $\gamma$ is of the order of ratio of the effective force acting on the electron $f \sim |e|E_{eff}$ and the average force of interaction of neighboring electrons in plasma $f_{el} \sim e^2 n_e^{2/3} \sim |e|E_{env}$ with the robust (but with possible future refinement) value $E_{env}$, where $n_e$ is electron density. In this case, with replacing $E_{env}$ by $E_{eff}$ where appropriate, all evaluations results of this work remain the same. Further developments of this work see ArXiv.org/physics.gen-ph/arXiv:1007.4377 .

Current theoretical models are extremely complicated, with the quantum exchange forces and superconductivity, vortex torus and magnetic traps, erosion-cluster-electrostatic accumulation of opposite charges, nuclear reactions, etc., which are indicated, for example, in recent works [5 - 7].

In this paper I present very simple considerations and estimates of parameters of BL showing the possibility of an elementary explanation the nature of BL with electric interactions, and also found a correlation between the size, temperature, voltage, direction of electric field lines with estimate of surface tension coefficient of BL (which is, as it is well known, very small).

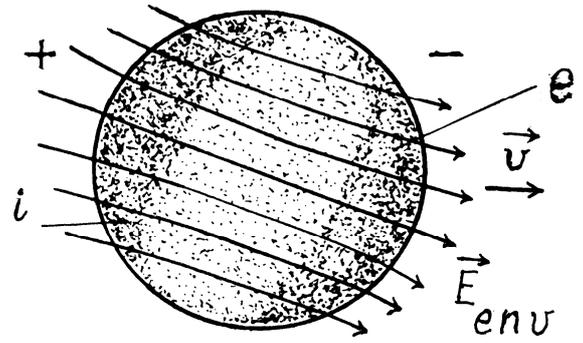

*Fig.1. Gas-plasma dipole*

It is assumed that BL consists of a partially ionized rarefied air with an internal pressure of the ambient medium $p \sim 1at$, with a dipole charge distribution produced by the external atmospheric electric field $E_{env}$ with an almost horizontal (sometimes vertical) vector field, the often movement of ball lightning near the Earth's surface which is usually charged positively (Fig.1).

Movement of spheroid-like gas-dipole BL is due to the asymmetry in relation to the movement of electrons and ions of BL near the anterior and posterior borders of BL in atmospheric electric field, so that the repulsion of the ions by electric field to the outside at the rear of the dipole is compensated by oppositely directed force of the reverse repulsion of ions by the cold dense atmospheric air. Thus, the ions move forward succeeding electrons in the rarefied air plasma of BL after their drift backward in external electric field and subsequent reverse recoil (cf. the discussion and estimates in [11]). Electrons are pulling a load of positive ions forward in the direction of the electric field strength with equal translation accelerations and velocities of overlapping electron and ion clouds in the final stage of formation and preserving the moving dipole.

Because much of Archimedean lift force acting on BL is approximately compensated by the vertical component of atmospheric electric field, BL can move along the horizontal component of the field vector, i.e. at an angle to the lines of force of electric field with the inclined axis of the dipole.

In turn, the slope of the electric field vector with respect to the horizontal surface of the Earth is due to a natural slope of the lines of force between the removing storm cloud and the local Earth's surface just after the last thunderstorm.

## 2. Integrity of BL

The necessary condition essential for BL as a holistic object is equal acceleration and velocity of general forward movement of ions and electrons. The preliminary the simplest ideas, as it can be assumed, are that the plasma BL (fireball, FB) has a common thermodynamic temperature, respectively, the neutral atoms and molecules may be partially captured by collisions with ions and to move with the same translational velocity. In addition, despite the usually somewhat asymmetrical oval or pear-like shape of BL, we can describe it for approximately as a slightly elongated ellipse or a sphere with a radius $r_{FB}$ with a predominance of electrons in the front part along the direction of motion and ions - in the back of the fireball (the approximate structure of dipole gas plasma).

In the simplest model one assumes that the force acting on the ions by the atmospheric electric field is compensated by resistance of the outer dense atmosphere for runaway ions in the back part of BL (the equivalent of repulsion (recoil) of the ions into the BL for their small mean free path in the outer atmosphere, as compared with the typical path of the electrons in front of the BL (cf. the later estimates in [11])).

In addition, more rigorous analysis must take into account the vertical component of the atmospheric electric field, compensating for Archimedean lift, with the turn of the dipole axis at a constant angle with relation to the lines of force and trajectory of BL, which can lead to some restriction of the minimum tension of electric field that supports the existence of BL. Acceleration of the ion cloud of BL plasma together with a part of neutral particles (neglecting the resistance of the ambient air to BL as an integrity, see below) is due to electrostatic force of the dipole interaction of ions with electrons of BL:

$$a_+ = \frac{e^2 N_e^2}{d^2 m_{av} V_{FB} n_L \frac{T_{env}}{T_{FB}}} = \frac{e^2 \left(n_L \frac{T_{env}}{T_{FB}} x_e V_{FB}\right)^2}{d^2 m_{av} V_{FB} n_L \frac{T_{env}}{T_{FB}}}, \quad (1)$$

where $T_{env} = 293\,\text{K}$; $T_{FB}$ is temperature of BL; $n_L = 2.5 \times 10^{19}\,\text{cm}^{-3}$ is Loschmidt number, which corresponds to 20 °C; $e$ is electron charge; $N_e \simeq N_i$ is number of electrons in the volume of BL (up to $T_{FB} \sim 12\,000\,\text{K}$ air plasma is dominated by singly charged ions [8] ),

$$V_{FB} = \frac{4}{3} \pi r_{FB}^3;\quad d \simeq 2 r_{FB}; \quad (2)$$

$d$ is effective length of the dipole, $r_{FB}$ is effective radius of the BL, with an average mass of heavy particles (atoms, molecules, ions) $m_{av}$ depending on $T_{FB}$; $x_e$ is molar fraction of electrons [8].

The acceleration of the electron cloud by an external electric field is equal

$$a_- = \frac{|e| N_e E_{env}}{N_e m_e}, \quad (3)$$

where $E_{env}$ is external electric field.

A necessary condition for the integrity of the charged part of the BL including both charged and neutral particles

$$a_+ = a_- \quad (4)$$

leads to the relation (5)

Table 1[*]

*BL parameters according to Eq. (5)*

| $T_{FB}$, K | $T_{env}/T_{FB}$ | $x_e$ | $m_{av}$, g | $E_{env}$, V/cm | $\rho_{FB}/\rho_0$ |
|---|---|---|---|---|---|
| 6000 | $4.9 \times 10^{-2}$ | $1.7 \times 10^{-4}$ | $3.7 \times 10^{-23}$ | $1.3 \times 10^{0}$ | $3.7 \times 10^{-2}$ |
| 7000 | $4.2 \times 10^{-2}$ | $6.0 \times 10^{-4}$ | $3.0 \times 10^{-23}$ | $1.7 \times 10^{1}$ | $2.6 \times 10^{-2}$ |
| 8000 | $3.7 \times 10^{-2}$ | $2.4 \times 10^{-3}$ | $2.6 \times 10^{-23}$ | $2.8 \times 10^{2}$ | $2.0 \times 10^{-2}$ |
| 8500 | $3.5 \times 10^{-2}$ | $4.6 \times 10^{-3}$ | $2.5 \times 10^{-23}$ | $1.00 \times 10^{3}$ | $1.8 \times 10^{-2}$ |
| 10 000 | $2.9 \times 10^{-2}$ | $2.3 \times 10^{-2}$ | $2.4 \times 10^{-23}$ | $1.74 \times 10^{4}$ | $1.3 \times 10^{-2}$ |

[*] $r_{FB} = 10\,\text{cm}$; $p_{FB} = 1\,\text{at}$; $\rho_0 = 1.2 \times 10^{-3}\,\text{g}/\text{cm}^3$ is atmospheric air density.

$$r_{FB} = \frac{3}{\pi} \frac{m_{av}}{m_e |e| x_e^2 n_L \dfrac{T_{env}}{T_{FB}}} \cdot E_{env}, \qquad (5)$$

where values of the external electric field $E_{env}$ in which is BL can range from several tens V/cm to several kV/cm at temperatures of ~(8000 - 9000) K (see Tables (1, 2)) (see also the later proof in [11]).[4] Note that the derivation of equation (5) neglects the resistance of the medium, which affects the BL as a whole. Note also that when you turn on braking $a_{brake}$ with keeping the integrity and the observed constant velocity we obtain $a_+ - a_{brake} = a_- - a_{brake} = 0$. In this context, the condition (5) corresponds to equal electrostatic forces acting correspondingly on electrons and ions in the uniform translational motion of BL.

The applicability of equations (1) - (5) communicates with the idea of BL (confirmed by further evaluations, see Tables 3-6), as a dipole with a substantial separation of charges (such as dumbbell) with a very tight electrostatic coupling between them (in the jumper between them) due to the strong electric field (see also Table 7). Thus, resistance to translational movement is determined by forces acting not on positive and negative charges apart, but on the dipole "gas-dumbbell" as a whole.

The existence of BL can be supported by an extended electric field with tension from several tens and hundreds to thousands of V/cm, while the emergence of a linear lightning needs rather less than tens of kV/cm (usually < (3 - 4) kV/cm for large distances ~ 1 km), breakdown voltage in normal air depending on the distance is $\lesssim 30$ kV/cm (but see also footnote 3 on p. 1).

*Table 2*
*Values $E_{env}$, V/cm according to Eq. (5)*

| $T_{FB}$, K \ $r_{FB}$, cm | 5 | 10 | 15 |
|---|---|---|---|
| 6000 | 0.65 | 1.3 | 2.0 |
| 7000 | 8.50 | 17.0 | 25.5 |
| 8000 | 140 | 280 | 421 |
| 8500 | 500 | 1000 | $1.50 \times 10^3$ |
| 10 000 | $8.7 \times 10^3$ | $1.74 \times 10^4$ | $2.6 \times 10^4$ |

### 3. The traveling speed of BL and its contained energy

The speed $v_{FB}$ of BL is determined by the balance between the inflow of energy per unit time $W_+ = |e| N_e E_{env} v_{FB}$, radiation losses $\varepsilon \sigma S T_{FB}^4$, $S = 4\pi r_{FB}^2$, and BL braking resistance of the surrounding atmosphere. Here $\varepsilon$ is emissivity of the BL which depends on temperature, size and geometric shape of the emitting object, $\sigma$ is Stefan-Boltzmann constant. Later, as a first approximation, it is adopted value $\varepsilon$ calculated in the center of a hemispherical base with a radius $r_{FB}$ [9]. This radiation is essentially non-Planckian, what leads to the possibility of different colors and luminosity of BL.

The movement solid body resistance is defined by Stokes formula $F_{res} = 4\pi \eta v_{FB} R_e$, where $\eta = 1.72 \times 10^{-4}$ g/(cm·s) is dynamic viscosity of air, and $R_e = \rho_{env} v_{FB} r_{FB} / \eta$ is Reynolds number.

---

[4] For account for BL plasma polarization degree see the footnote 3 on page 1.

If $r_{FB} = 10$ cm, $v_{FB} = 1$ m/s, then we obtain $R_e = 1.40 \times 10^3$ and the power losses $W_{res} = F_{res}v_{FB} = 7.6$ W (only assuming that the gas-plasma BL behaves like a solid body with severely fixed hard boundaries!).

At more speeds the resistance force is characterized by the relation $F_{res} = C_x \pi r_{FB}^2 \rho_{FB} v_{FB}^2 / 2$, where $C_x$ is the well-known (tabulated) resistance factor $C_x \leq 0.4$, that is at a speed of 1 m/s we obtain $W_{res} = v_{FB} F_{res} \sim 4 \times 10^{-3}$ W, and if we have $v_{FB} \sim 100$ m/s then $W_{res} < 40$ W. Therefore in future we can neglect these losses compared with radiation losses.

For gaseous BL it appears to be also possible lateral losses due to ambipolar diffusion with the coefficient $D_a \sim D_i \cdot (1 + T_e/T_i)$, which in this highly simplified model are not accounted for (cf. [11]).

The traveling speeds of BL calculated according to the formula

$$v_{FB} = \frac{\varepsilon \sigma T_{FB}^4 S}{|e| N_e E_{env}} \qquad (6)$$

are presented in Table 3.

The calculations in Table 3 assumed complete capture of neutral molecules by BL. The radiation power

*Table 3*

*Thermodynamical equilibrium calculated parameters of BL*

| $r_{FB}$, cm | 5 | | | | | 10 | | | | | 15 | | | | |
|---|---|---|---|---|---|---|---|---|---|---|---|---|---|---|---|
| $T_{FB}$, K | $\varepsilon \times 10^3$ | $\varepsilon \sigma T_{FB}^4 S$ W $\times 10^{-3}$ | $|e| N_e E_{env}$ kgf $\times 10^{-2}$ | $v_{FB}$ m/s | | $\varepsilon \times 10^3$ | $\varepsilon \sigma T_{FB}^4 S$ W $\times 10^{-4}$ | $|e| N_e E_{env}$ kgf $\times 10^{-3}$ | $v_{FB}$ m/s | | $\varepsilon \times 10^3$ | $\varepsilon \sigma T_{FB}^4 S$ W $\times 10^{-5}$ | $|e| N_e E_{env}$ kgf $\times 10^{-4}$ | $v_{FB}$ m/s | |
| 6000 | 0.5 | 1.2 | $1.2 \times 10^{-3}$ | 1000 | | 1.1 | 1.0 | 0.001 | 1000 | | 1.6 | 0.33 | 0.0010 | 330 | |
| 7000 | 0.7 | 3.0 | $7.6 \times 10^{-2}$ | 39 | | 1.3 | 2.2 | 0.070 | 31 | | 1.9 | 0.73 | 0.037 | 19.7 | |
| 8000 | 1.1 | 8.0 | 2.6 | 4.0 | | 1.6 | 4.7 | 4.2 | 1.2 | | 2.4 | 1.6 | 2.1 | 0.76 | |
| 8500 | 1.8 | 17 | 17 | 1.0 | | 2.3 | 8.6 | 27 | 0.31 | | 3.4 | 2.9 | 13.4 | 0.22 | |

is implausibly large, and the values of the acting BL electrostatic force which are calculated assuming the total dipole charge separation, seem fantastic.

It may be supposed also distribution of momentum transfer to all electrons from an external electric field due to compensation of the opposite momentum transferred to ions on account for their repulsions by cold dense air molecules near the back border of BL. It might be noted also that for realistic variants we have $e^2/l_{av}^2 \lesssim |e| E_{env}$ where $l_{av} \sim 1/n_e^{1/3}$ (see Table 7). Thus, on the microscopic level, the Coulomb interactions between particles can be much more then their interactions with external electric field (for realistic account of polarization see footnote 3).

More plausible results were obtained following both the assumption of complete capture of neutral molecules by the fireball and substantial thermodynamic non-equilibrium BL with $T_{env} \ll T_m \ll T_e$ and introduction non-equilibrium parameter $\alpha$ (see further Table 6).

Upon the termination of inflow energy (with the disappearance of the external ambient field or stop before the obstacle for BL) for a short time part of the energy is going away in the form of radiation, the remaining energy of the BL can be released with the explosion. For example, we estimate the maximum energy of the BL after the partial release of radiation for the case of $T_{FB} = 8000\text{K}$; $r_{FB} = 5\text{cm}$.

1. The energy of recombination radiation in the line spectrum $E_{rec} \sim I N_e \simeq 2.7\text{ J}$, where $I \sim 15\text{ eV}$ is ionization potential.

2. Electrostatic energy $E_{el} \sim |e| N_e \cdot 2 r_{FB} E_{env} = (e N_e)^2 / 2 r_{FB} \simeq 250\text{ J}$.

3. Heat energy $E_h \sim n_L \dfrac{T_{env}}{T_{FB}} \cdot \dfrac{7}{2} \kappa (T_{FB} - T_{env}) \cdot V_{FB} \simeq 174\text{ J}$.

4. Energy of the BL cavity filling with ambient air is $E_{fill} \sim p_{env} \cdot V_{FB} \simeq 53\text{ J}$.

Thus, the maximum energy of the explosion at density $E/V \gtrsim 10^6\text{ J/m}^3$ can reach up $(10^3 - 10^4)\text{ J}$ and over (cf. [11]), depending on radius, what is sufficient for the observed not only local damage, such as brick chimneys, window glass, household items, but even the major destruction, until the partly destruction of premises.

Variability of forms and manifestations of BL is determined by variation of free parameters of BL. Formal assignment of initial parameters is leading to the settlement both real and unrealizable regimes of BL. Tables (1 – 3) dramatically demonstrate the presence of an extremely narrow parametric field of regimes BL with a very sharp changes in the calculated parameters of BL on the boundaries of this parametric area. Thus, the most frequently observed size of the BL about $r_{FB} \sim 5\text{ cm}$ would correspond to the calculated velocity of movement of BL up to tens meters per second, typical calculated temperature of BL in a narrow range is close to 8000 K and there is often considerable destructive capability of BL explosion at the end of its life. At the same time the huge power of radiation does not match the observations and makes doubtful the possibility of the existence of such regimes.

Options BL are calculated from two equations of balance of acceleration of ions and electrons and balance of energy. However, great initial acceleration of free electrons in an electric field appears at first glance clearly not appropriate with respect to small resultant velocity of motion of BL. It can be seen as an almost instantaneous saturation at the end of the initial stage of transient accelerations and velocities of electrons and ions from the beginning of the second stage of movement, respectively, the energy balance follows zero accelerations and equal velocities of the ions and electrons. The ground for the application of equations (1) - (5) is the emergence of resistance to the translation movement of the BL as a whole, i.e. with the conservation of equations (1) - (5), after a very short transition period followed by maintaining a constant velocity and temperature. This resistance to movement is due mainly to radiation losses with transformation of this to mechanical resistance.

Fast randomization of electrons [11] leads to an increase in the electron temperature of the electron cloud with a sharp increase in the rate of inelastic processes of ionization, recombination, excitation of electronic and vibrational states of molecules, what, besides of the Coulomb attraction of ions, leads to a rapid redistribution of the momentum being received by electrons throughout all the mass of BL [11].

Experimental reproduction and measurements of BL may not seem to be overly expensive. It appears possible a real use, for example, focused microwave radiation (microwave or other devices) to heat the seed source of BL with the rarefied air plasma at temperatures up to $T_e \sim 10\,000\text{ K}$ at a pressure of $p = 1$ at ($r_{FB} \sim (5-10)\text{ cm}$) and constructing the extended system of electrodes to create an extended (not more than a few meters) permanent bulk electric field of the order few kV/cm with a some slope of lines in relation to the horizontal plane near the surface of the Earth. The system of the trajectory field should include vertical and horizontal plane electrodes of large size. The required power of the microwave discharge might be not less than $(20 - 50) \times 10^3\text{ W}$ at tension of the trajectory field up to $\gtrsim (0.1 - 1.0)$ kV/cm (cf. Tables 2 and 3 options with a $T_{FB} \sim (8000 - 8500)\text{ K}$, $r_{FB} \sim 5\text{cm}$).

Impose on the spherical surface of BL with radius $R$ imaginary grid of small rectangular cells of side length equal to the average distance between charged particles $l_{cell} = 1/(2n_e)^{1/3}$, with the cell square $S_{cell} = l_{cell}^2$ per particle, each charged particle in the cell node is attracted, in the first order, by four oppositely charged particles in the nearest cell sites. Accordingly, there is additional force directed toward the center of the sphere, and the additional pressure due to the curvature of the spherical shell equals, as it is easy to determine

$$\Delta p = \frac{f_{cell}}{S_{cell}} = \frac{2e^2}{l_{cell}R} \cdot \frac{1}{S_{cell}} = \frac{2e^2}{l_{cell}^3 R} = \frac{4n_e e^2}{R} = \frac{2\sigma_s}{R}, \qquad \sigma_s \simeq 2e^2 n_e, \tag{7}$$

where $\sigma_s$ is surface tension coefficient. Since the denominator contains the third power of $l_{cell}$, the following orders of the nearest particles at a rough estimate can be ignored (although they can be easily incorporated as an amendment). In the particular case of $T_{FB} = 8500$ K we obtain $\sigma_s \sim 2 \times 10^{-3}$ mN/m, which is much smaller than the surface tension of liquids such as ether (17 mN/m), alcohol (23 mN/m), acetone (24 mN/m) or water 73 mN/m. With increasing temperature the concentration of charged particles increases and, consequently, $l_{cell}$ decreases with a corresponding increase $\sigma_s$. In this way BL cloud, it would seem, can not be sharply distinguished from the surrounding air and can not behave in some cases like a fluid, penetrating through the holes if they are not too small. But it is shown in [11], that due to ambipolar diffusion BL has significant positive charge which keeps BL from dispersion and is analogue of surface tension.

Surface tension has no effect on neutral particles which are more or less entrained only by collisions with charged particles with ionization and recharging. Thus, it can lead to rapid movement mainly of the charged particles with an inactive immobile neutral component of the BL, however leaving a hot trail after passing BL and possibly resulting in convective rotation of BL [10]. At the same time at a characteristic relative mean velocity of neutral particles

$$v_{av} = \sqrt{\frac{3}{2} \cdot \frac{kT_{FB}}{m_{av}/2}} \sim 3.45 \times 10^5 \text{cm/s} \gg v_{FB}(8500 \text{ K})$$

and the cross section $\sigma_{ex} \sim (10^{-14} \div 10^{-15})$ cm$^2$ of charge exchange at collisions of neutral particles with ions we obtain for the mean free path of neutral particles with recharging

$$l_{ex} = \frac{1}{\sigma_{ex} n_i} \sim (0.025 \div 0.25) \text{ cm} \ll r_{FB}.$$

The hypothetical limiting case, which might appear more real, of an immobile part of BL neutrals can be evaluated according to formulas (1), (5) with replacing $x_e^2 \to x_e$, the temperature of heavy particles $T_m < T_e$ and the choice of $x_e, \varepsilon$ correspondingly to temperature $T_e$. For example, choosing arbitrary values

$$T_m = 3000 \text{ K}; \quad T_e = 6000 \text{ K}; \quad r_{FB} = 5 \text{ cm, and } m_{av} \simeq 3.7 \times 10^{-23} \text{g};$$
$$n_L = 2.5 \times 10^{19} \text{cm}^{-3}; \quad T_{env} = 293 \text{ K}; \quad x_e = 1.7 \times 10^{-4} \tag{8}$$

we obtain the following estimate:

$$E_{env} = \frac{\pi}{3} \cdot \frac{m_e}{m_{av}} \cdot |e| x_e n_L r_{FB} \cdot \frac{T_{env}}{T_m} \simeq 7.7 \text{ kV/cm (!!)};$$
$$v_{FB} = \frac{3\varepsilon \sigma T_e^4}{eE_{env} r_{FB}} \cdot \left( n_L x_e \frac{T_{env}}{T_m} \right)^{-1} \simeq 0.01 \text{ m/s}; \quad |e|E_{env} N_e \simeq 2.8 \times 10^3 \text{kgf (!!)}. \tag{9}$$

Numerous eyewitness accounts speak in favor of the non-equilibrium "cold" BL with radiation power of the order of power the light lamp ~ (100 - 200) W, in contrast to the previously cited results of Table 3.

*Table 4*

*Variants of BL without confined neutral particles*

| $T \equiv T_{FB} =$ $= T_m$, K | $x_e$ $\times 10^5$ | $n_e$ cm$^{-3}$ $\times 10^{-13}$ | $N_e$ $\times 10^{-16}$ | $\varepsilon$ $\times 10^4$ | $\varepsilon \sigma T^4 S$ W | $E_{env}$ V/cm | $|e| N_e E_{env}$ kgf | $v$ m/s |
|---|---|---|---|---|---|---|---|---|
| 3000 | 0.32 | 0.79 | 0.42 | ~1.1 (?) | 15.7 | 116 | 0.8 | 2.0 |
| 4000 | 1.37 | 2.5 | 1.3 | 1.7 | 78 | 409 | 8.7 | 0.89 |
| 5000 | 4.7 | 6.9 | 3.6 | 3.6 | 396 | 1210 | 71 | 0.56 |
| 6000 | 17 | 42 | 22 | 5 | 1160 | 6200 | 220 | 0.53 |

*Table 5*[**]

*BL parameters without confined neutral particles but with account for non equilibrium parameter $\alpha > 1$*

| $T_e$, K | $T_m$, K | $\varepsilon$ $\times 10^3$ | $x_e$ | $\alpha$ | $W_\varepsilon$ W | $\beta$ | $n_e^{(FB)} = n_e/\alpha$, cm$^{-3}$ $\times 10^{-15}$ | $E_{env}$ V/cm | $F_{FB}$ kgf | $v_{FB}$ m/s |
|---|---|---|---|---|---|---|---|---|---|---|
| 4000 | 400 | 0.17 | $1.4 \times 10^{-5}$ | 1.5 | 34.7 | 0.21 | 0.167 | 0.11 | 0.016 | 212 |
| 5000 | 400 | 0.36 | $4.7 \times 10^{-5}$ | 1.5 | 176 | 0.72 | 0.57 | 0.51 | 0.25 | 71.5 |
| 7000 | 400 | 0.70 | $6.0 \times 10^{-4}$ | 4.5 | 150 | 4.1 | 3.29 | 8.42 | 23.5 | 0.90 |
| 8000 | 400 | 1.1 | $2.4 \times 10^{-3}$ | 6.3 | 201 | 8.7 | 6.95 | 36.0 | 224 | 0.090 |
| 8500 | 400 | 1.8 | $4.6 \times 10^{-3}$ | 8.4 | 236 | 25 | 10.1 | 162 | 1390 | $1.73 \times 10^{-3}$ |
| 10 000 | 400 | 5.0 | $2.3 \times 10^{-2}$ | 20 | 222 | 26 | 21.0 | 343 | 6660 | $3.4 \times 10^{-3}$ |
| 8500 | 3000 | 1.8 | $4.6 \times 10^{-3}$ | 9.1 | 200 | 1.16 | 0.93 | 9.7 | 7.7 | 2.53 |
| 9000 | 3000 | 2.3 | $8.3 \times 10^{-3}$ | 11.6 | 200 | 2.15 | 1.72 | 20.4 | 30.0 | 0.68 |
| 10 000 | 3000 | 5.0 | $2.3 \times 10^{-2}$ | 18.9 | 250 | 3.71 | 2.97 | 52.8 | 99.4 | 0.26 |
| 12 000 | 3000 | ~15 | $1.04 \times 10^{-1}$ | 52.6 | 200 | 6.0 | 4.85 | 123 | $5.6 \times 10^2$ | 0.036 |

[**] $r_{FB} = 5$ cm; $\beta \simeq 2.5 \times 10^{-15} r_{FB} n_i / \gamma_{coll}$, $\gamma_{coll} \sim 10$, $m_{av} \equiv m_{av}(T_m)$, $F_{FB} = |e| N_e E_{env}$; about non-equilibrium parameter $\alpha$ see below.

Preliminary assessments (8), (9) can be regarded as a hint for the possibility of simultaneous, along with exotic high temperature (possibly less stable, or due to differences between the initial seeding level), also options with a substantially non-equilibrium plasma and immobile gas component of neutral atoms and molecules, but within the purely electric dipole plasma model with external sources of energy from atmospheric electric field (that is, the sliding of the charged components at fixed neutral ones).

For the limiting case when the neutral particles do not entrain, but only are heated by the charged particles, leaving behind a trail of heated air, the results of calculations at $r_{FB}=5\,\text{cm}$ are presented in Table 4.

If $\xi$ is relative amount of neutral molecules, entrained by the BL, and $T_e \neq T_m$, then Eq. (5) for $E_{env}$ with $x_e^2 \to x_e$ can be rewritten as

$$E_{env} \simeq \frac{\pi}{3}\frac{m_e}{m_{av}}|e|r_{FB}\frac{n_e^2}{n_e+\xi(n_m-n_e)}. \qquad (10)$$

Accordingly to this, neglecting the losses due to excitation of electron and vibrational states of molecules,

$$v_{FB} \simeq \frac{\varepsilon \sigma T_{FB}^4 S}{-\frac{7}{2}k(T_m - T_{env})n_m \cdot \pi r_{FB}^2 \cdot (1-\xi) + (\lambda+\mu)|e|N_e E_{env}}, \qquad (11)$$

where $\lambda$ is the proportion of energy expended for heating of the neutral molecules, which determines their temperature $T_m$, and $\mu$ is the dominant share of energy, ultimately spent on the radiation and other losses,

$$\xi = 1 - e^{-\beta}, \qquad (12)$$

and $\beta$ is proportional to the product cross section of collisions of neutral molecule with ions (including recharging),

$$\beta \sim \sigma_c n_i r_{FB}/\gamma_{coll}, \quad n_i = n_e, \qquad (13)$$

$\gamma_{coll}$ is the average number of collisions of neutral molecule with ions, necessary for the transfer of translational momentum of the ion.

In the first approach we can take into account non-equilibrium $T_{FB} = T_e > T_m$ with a decrease $\varepsilon$ and $x_e$ correspondingly: $\varepsilon \to \varepsilon/\alpha^2$ и $x_e \to x_e/\alpha$, $n_e = x_e n_m = x_e n_L \cdot (293/T_m)$; $\alpha \geq 1$, because the rate of recombination is proportional to $n_e^2$, and the ionization rate is proportional to $n_e$.

"Cold" BL should also be consistent with the rather lower temperature $T_m < T_{FB}$ condition which imposes also constraints on the minimal values of $E_{env}$ necessary to support electrical discharge of BL.

In this case for the applicability of the results Table 4 there should be fulfilled also the obvious additional constraint conditions (14), (15)

$$\xi(n_m - n_e) \ll n_e, \qquad (14)$$

$$\frac{7}{2}k(T_m - T_{env})n_m \cdot \pi r_{FB}^2 \cdot (1-\xi) \ll |e|N_e E_{env}. \qquad (15)$$

These conditions, however, are not satisfied with the accepted in Table 4 values $\beta$ and $T_m = T_e$, and therefore further we considered some options for substantially non-equilibrium plasma with a lower temperature $T_m$ and resulting large values $\beta$ (i.e. with full capture of neutral molecules by fireball) with a calculation by formulas (10) - (13) at $r_{FB} = 5$ and 10 cm, see Tables 5, 6.

It might be noted also that at electrostatic force $|e|E_{env}$ can be of one - two orders of magnitude less of

*Table 6*$^{***}$

*High temperature thermodynamically non-equilibrium variants of BL*

| $T_e$, K | $T_m$, K | $\varepsilon \times 10^3$ | $x_e$ | $\alpha$ | $W$, W | $\beta$ | $F_{FB}$, kgf | $v_{FB}$, m/s | $n_e^{(FB)} \times 10^{-15}$, cm$^{-3}$ | $E_{env}$, V/cm |
|---|---|---|---|---|---|---|---|---|---|---|
| 10 000 | 1000 | 6.8 | 0.0233 | 49.3 | 200 | 8.67 | 1100 | 0.0185 | 3.47 | 46.6 |
| 12 000 | 1000 | 25 | 0.104 | 136 | 200 | 14 | 4670 | 0.0044 | 5.61 | 122 |
| 12 000 | 3000 | 25 | 0.104 | 136 | 200 | 4.67 | 494 | 0.0413 | 1.87 | 38.6 |
| 12 000 | 6000 | 25 | 0.104 | 136 | 200 | 2.33 | 186 | 0.11 | 0.93 | 29.1 |

*** $r_{FB} = 10$ cm, $\beta = 2.5 \times 10^{-15} r_{FB} n_i / \gamma$, $\gamma \sim 10$, $m_{av} = m_{av}(T_m)$.

*Table 7*$^{****}$

*Account for polarization with increasing atmospheric electric field $E_{env}$ by a factor $\gamma_p^{-1}$*

| $T_e$, K | 8000 | 10 000 | 8000 | 8500 | 10 000 | 12 000 |
|---|---|---|---|---|---|---|
| $T_m$, K | 8000 | 10 000 | 400 | 400 | 400 | 3000 |
| $\alpha$ | 1.0 | 1.0 | 6.3 | 8.4 | 20 | 52.6 |
| $\gamma_p E_{env}$, V/cm | 140 | $8.7 \times 10^3$ | 36 | 162 | 343 | 123 |
| $e^2/l_{av}^2$, dyn | $3.9 \times 10^{-9}$ | $1.4 \times 10^{-8}$ | $8.3 \times 10^{-9}$ | $1.06 \times 10^{-8}$ | $1.75 \times 10^{-8}$ | $6.35 \times 10^{-9}$ |
| $|e|E_{env}$, dyn | $2.2 \times 10^{-10}$ | $1.4 \times 10^{-8}$ | $5.8 \times 10^{-11}$ | $2.6 \times 10^{-10}$ | $5.5 \times 10^{-10}$ | $2.0 \times 10^{-10}$ |

**** With using Tables 2 and 5; $r_{FB} = 5$ cm.

the force of Coulomb interaction between the electron and ion at a distance $l_{av} \sim [(n_e/\alpha)]^{1/3}$ (see Table 7).

In these cases in accordance with the consideration in the work [11] the tension of the atmospheric electric field $E_{env}$ should be increased at least by an order of value with a factor $\gamma_p^{-1} \gtrsim 10,$ and in all formulas $E_{env}$ should be replaced for $\gamma_p E_{env}$, $\gamma_p = E_{env}/|e|(n_e/\alpha)^{2/3}$ (cf. footnote 3 on p.1).

It is assumed that between oppositely charged particles and neutrals occurs uniform distribution of the momentum transferred to all electrons by an external electric field due to compensation of the opposite momentum transferred to ions by this electric field at their repulsion from molecules near the back border of BL. The resistance to BL forward movement is created by recoil at $i - e$ recombinations in the front site of fireball (FB).

With incomplete compensation in equation (1) of opposite force of recoil momentum $|e|N_i E_{env}$ acting on ions on the back of BL, while maintaining the size of the dipole, the term $\tau |e| N_i E_{env}/M_{FB}$ ought to be added, where $M_{FB}$ is mass of BL and $\tau$ is the share of uncompensated acceleration. Then equation (5) becomes

$$E_{env} = \frac{\pi}{3} \cdot \frac{m_e}{m_{av}} \cdot \frac{|e|n_m x_e^2 r_{FB}}{1 - \frac{m_e}{m_{av}} \tau x_e}, \quad 0 < \tau < 1, \qquad (16)$$

That is the field tension can increase, but very little. It should mean some weakening of the wake recoil trace.

The recombination radiation that is supported by supplying the energy from the outside electric field regulates the rate of recombination in the BL and keeps it from collapsing. The internal field of charges interaction keeps BL from the free-diffusion dissipation. The smallness of the forced flow of runaway ions due to their low backward mobility in an external electric field [11] determines its forward movement and energy supply from an external field.

### 4. Conclusion

At presence of a large number of often rather exotic hypotheses to explain the appearing mysterious properties of BL, we provided basic, very simple examination of fireball associated with the presence in atmosphere the inclined lines of weakened electric field of going away thunderstorm cloud, which is generating at some point the dipole plasma formation BL moving horizontally over the Earth's surface with the energy supplying by atmospheric electric field. In the limiting case of thermodynamic equilibrium $T_e = T_m$ ("hot" version) there are estimated parameters of BL with typical sizes, temperature and them corresponding speed and radiation power of BL. There is considered further the rather most realistic case of highly thermodynamically non-equilibrium BL plasma ("cold" version). There is found a linear dependence of the size $r_{FB}$ of BL on the intensity of atmospheric electric field. We estimate the energy of FB released in explosion at the end of BL existence (see also estimates in [11]).

These calculations are made taking into account the horizontal component of the electric field, neglecting the small vertical component, which produces a force, compensating Archimedean lift force acting on BL. Thus, the dipole BL can move horizontally relative to the surface of the Earth with the axis of the dipole, tilted with respect to the trajectory and direction of electric field lines. Integrity of BL including neutral molecules is determined only slightly by the presence of surface tension forces, but mainly by internal attraction electrical forces acting on the electron charge component of the positively charged BL and a substantial role of collisions of neutral particles and ions with their recharging and relatively low losses of ambipolar diffusion at high electron temperatures $T_e$ (cf. [11]).

In the future one should perform refinements, taking into account compensation for BL particles escaping (ions in the rear of BL and neutral molecules into the surrounding dense air), value of recoil momentum which, in fact, provides the asymmetry of the dipole BL motion under the electric field due to the effect of repulsion and return of the ions back into the BL plasma. Although the significant evaluations are performed, substantial new updates need to be made (stability analysis of BL and random variability of ambient atmospheric electrical field, a more accurate account of the radiation losses and estimations of the minimal electrical field supporting BL, relation of non-equilibrium electron and ion temperatures, as well as the incomplete compensation of the recoil forces acting on the ions by the atmospheric electric field due to the mobility ions in the back of the BL, etc.), the main first specified value of this work is demonstration in principle the possibility of simultaneous explaining of many of the properties and behavior of BL with purely electrical processes partly characteristic of electric discharges in rarefied gases, with the possibility of experimental modeling BL.

Performed above estimates confirm the possibility of existence, perhaps, more stable and consistent with observations, the "cold" BL ($T_i = T_m \sim 1000 - 3000$ K) with a strong non-equilibrium plasma, observed relatively low but crucial non-equilibrium radiation power and probably almost captured neutral component (without significant slipping of BL through the neutral medium, that is without slipping the charged component of the ions and electrons as a whole relative to not confined neutrals). It is assumed that the idea of the primarily non-equilibrium, defined by the parameters $T_e/T_m > 1$ and non-equilibrium parameter $\alpha$, weak ionized BL plasma with correspondingly small electron mole fraction $x_e$ and small blackness degree $\varepsilon$, will provide the most possible type of BL (moreover, in a subsequent paper [11] it is shown the impossibility of the existence of stable thermodynamic equilibrium BL).

The electrons receive energy and momentum from an external electric field, the last is then distributed through collisions in the entire FB as a whole (without discrimination of electrons and ions forming fireball moving in the external air environment).

In addition to the excess positive charge of BL as analogue of the surface tension the optimized charge geometry also plays some role, although the usual surface tension is vanishingly small. The electric energy acquired by BL in its motion in electric field is converted into radiation energy which is the main accessible to observation way to probe "hot" and "cold" BL versions.

The presence of two free parameters in the non-equilibrium model BL leads to hypothesis of the possible existence of two other different degenerate branches of the hot and cold type BL: (1) high temperature $T_e$ at high non-equilibrium parameter $\alpha$; (2) somewhat lower temperature $T_e$ with relatively small $\alpha$ and at only slightly different in both cases temperatures $T_m$. These solutions with almost the same $n_e$ but different $T_e$ would correspond to "chameleon" BL, for example, color variability in the course of its flight.

The experimental difficulties associated with the reproduction of BL need to create an enough extended trajectory of the extremely strong electric field and the initial (seed) low density discharge, might be focused microwave or other, with characteristic parameters of BL (see Tables 1 - 6).

Required seed perhaps might be a microwave discharge with a size of about (5 - 20) cm and temperatures up to $T_e \sim (8000-12000)$ K; $T_m \sim (1000-3000)$ K at atmospheric pressure, and construction of large electrode system to create the trajectory of electric field simulating the atmospheric tilted electric field of removing thundercloud.

The lifetime of BL can also particularly be caused by instability of the atmospheric electric field. The linear lightning might not be the cause of the huge power of BL, on the contrary, the last is merely an indicator of regions of strong atmospheric electric fields and may be trigger for the breakdown and the local lightning.

Surprising exotic features of the dipole model are its stability and relatively long lifetime which are inextricably linked to its forward movement with constant speed due to resistance by the dominant radiation losses.

Further discussion and new evidences with estimates of recoil momentum, movement resistance, stability and ambipolar diffusion losses are presented in [11].

Highly thermodynamically non-equilibrium dipole model of ball lightning well explains its basic properties: stability (see [11]), integrity, translational speed, moderate luminosity, it is characterized by observed low gas temperature of $T_m \sim 10^3$ K (with a high electron temperature of $T_e$ up to $\sim 10^4$ K), the ability to pass through intact glass, high specific energy content (see [11]) and electrical injuries, the usual absence of direct communication with the linear lightning. There are not ruled out the variability of geometric forms of BL and the ability to pass through the not too small holes.

## References

1. Singer S. The nature of Ball Lightning, Plenum, NY (1971).
2. McNally J. R., Jr. Preliminary Report on the Ball Ligtning. Oak-Ridge National Laboratory, Report n. ORNL 3938. 1966.
3. Rayle W. D. Ball Lightning Caracteristics. NASA Technical NOTE-D-3188. 1966.
4. Stakhanov I. P. On physical nature of Ball Lightning. Energoatomizdat, Moscow, 1985 (in Russian).
5. Bychkov V, L. Unipolar Ball Lightning Model. Proc. of the 8$^{th}$ International Symposium on Ball Lightning (ISBL 04), Chung-li, Taiwan. 2004.
6. Nikitin A. I. The Dynamic Capacitor Model of Ball Lightning. Proc. of the 6$^{th}$ International Symposium on Ball Lightning (ISBL 99). Antwerp. Belgium. 1999.
7. Vlasov A. N. Magnetohydrodynamical Model of plasma object capable to be generated at impact of ordinary Lightning. Proc. of the 10th International Symposium on Ball Lightning (ISBL 08) and Unconventional Plasmas (ISUP 08). Kaliningrad, Russia. 2008.
8. Predvoditelev A. C. *et al.* Tables of thermodynamic functions of air (for temperatures from 6000 K up to 12 000 K) and pressures from 0.001 up to 1000 at. M. "Наука". АН СССР (RAS USSR), 1959 (in Russian).
9. Avilova I. V., Biberman L. M., Vorobjev V. S. *et al.* Optical properties of hot air. M. "Наука", АН СССР (RAS USSR), 1970 (in Russian).
10. Domokos Tar. Observation of Lightning Ball (Ball Lightning): A new phenomenological description of the phenomenon. ArXiv.org/Physics/Plasma physics/arXiv:0910.0783.
11. Soshnikov V. N. Comments in support of the Dipole Model of Ball Lightning. Arxiv.org/plasma physics/arxiv:1007.4377.

**Acknowledgement.** I express my sincere gratitude to V. L. Bychkov (Head of the Department of Physical Electronics, Moscow State University) for constructive criticism and an indication of the existing vast literature.

## Addendum

Theoretical model of ball lightning (BL) has identified the main properties that do not contradict to its something ambiguous observable phenomenology. That is:

1. BL is a strong gas-plasma dipole with size ~ (10 - 20) cm, moving at varying speeds from few millimeters to tens of meters per second or more (cf. also [11]) under the influence of nearly horizontal tilted electric field of receding storm cloud.
2. BL is a compact heterogeneous formation composed as a whole from the charge of air plasma with an electron temperature $T_e \sim (8000 - 12\,000)$ K and retained neutral connected component at temperature $T_m \sim T_i < T_e$.
3. Properties of BL are defined by the fundamental role of light radiation in the energy balance of BL.
4. Plasma of BL is essentially thermodynamically non-equilibrium at temperatures $T_m \sim T_i$ rather less than $T_e$.
5. Movement of BL is tightly associated with the formation behind it recoil wake (jet) of cold or only slightly heated air through the law of conservation of momentum.
6. The strong thermodynamic non-equilibrium determines prevailing of the so-called "cold" BL with moderate light radiation.
7. In accordance with the previously well known estimates, BL has a very weak surface tension (but large excess positive charge, see [11]).
8. In considering the non-equilibrium modifications of BL-model, density of charged particles is constrained in a narrow range of densities of orders $n_e \sim (10^{14} - 10^{15})\,\mathrm{cm}^{-3}$ at the supporting model external electric field $E_{env} \sim \mathrm{kV/cm}$ (Tables 6 - 7).
9. The real diversity of BL can be caused not only by differences of $E_{env}$ but possibly different initial (seed) conditions of the subsequent stable BL.

10. Essentially dipole character of BL and availability of very weak surface tension and positive charge might lead to its eventual passage through intact glass and not too small openings.
11. The blurred outlines and sparking of BL may be due to running of a small part of excited metastable molecules not interacting with the electric field and recombining of run away charges.
12. The destructive energy $E_{est}$ of BL can be determined by the predominance of electrostatic dipole interaction and can reach in the given examples the values $E_{el} \sim (10^3 - 10^4)$ J and more, depending on radius ($\gtrsim 10^6 \text{ J}/\text{m}^3$).
13. Leaks of the excited molecules must be compensated by an additional increase of the speed $v_{FB}$ and/or $E_{env}$.
14. Lifetime of BL can substantially depend, besides of intrinsic and kinetic instability, also on the instability of the atmospheric electric field.

A rather more detailed analysis, evaluations and calculations are presented in the later revised and completed work [11].